\documentclass[11pt]{article}

\usepackage{bm}
\usepackage[normalem]{ulem}
\usepackage[utf8]{inputenc}
\usepackage[T1]{fontenc}
\usepackage{mathptmx}
\usepackage{etoolbox}
\usepackage{xcolor}
\usepackage{physics}

\usepackage[a4paper, margin=0.7in]{geometry}
\usepackage{authblk}
\usepackage{graphicx}
\usepackage{dcolumn}
\usepackage{amsmath}
\usepackage{amssymb}
\usepackage{textcomp}
\usepackage{verbatim}
\usepackage{float}
\usepackage[symbol]{footmisc}
\setfnsymbol{wiley}
\usepackage{hyperref}
\hypersetup{
    colorlinks=true,
    linkcolor=blue,
    filecolor=blue,
    urlcolor=blue,
    citecolor = blue,
    pdfauthor=author,
}

\usepackage{multirow}
\usepackage[T1]{fontenc}
\usepackage[normalem]{ulem}
\usepackage[numbers,sort&compress]{natbib}
\usepackage[utf8]{inputenc} 

\begin{document}

\title{Effects of surface roughness and top layer thickness on the performance of Fabry-Perot cavities and responsive open resonators based on distributed Bragg reflectors}%

\author[1]{K. Papatryfonos}
\author[1]{E.R. Cardozo de Oliveira}
\author[1]{N. D. Lanzillotti-Kimura\footnote{daniel.kimura@c2n.upsaclay.fr}}

\affil[1]{Université Paris-Saclay, CNRS, Centre de Nanosciences et de Nanotechnologies, 91120 Palaiseau, France}

\date{}
\maketitle
\begin{abstract}

Optical and acoustic resonators based on distributed Bragg reflectors (DBRs) hold significant potential across various domains, from lasers to quantum technologies. In ideal conditions with perfectly smooth interfaces and surfaces, the DBR resonator quality factor primarily depends on the number of DBR pairs and can be arbitrarily increased by adding more pairs. Here, we present a comprehensive analysis of the impact of top layer thickness variation and surface roughness on the performance of both Fabry-Perot and open-cavity resonators based on DBRs. Our findings illustrate that even a small, nanometer-scale surface roughness can appreciably reduce the quality factor of a given cavity. Moreover, it imposes a limitation on the maximum achievable quality factor, regardless of the number of DBR pairs. These effects hold direct relevance for practical applications, which we explore further through two case studies. In these instances, open nanoacoustic resonators serve as sensors for changes occurring in dielectric materials positioned on top of them. Our investigation underscores the importance of accounting for surface roughness in the design of both acoustic and optical DBR-based cavities, while also quantifying the critical significance of minimizing roughness during material growth and device fabrication processes.
\end{abstract}

\section{\label{sec:level1}Introduction}

Distributed Bragg reflector (DBR) devices based on semiconductor heterostructures are pivotal components in both fundamental and applied fields of photonics and nanophononics. A pair of DBRs enclosing an optical spacer constitutes a Fabry-Perot optical cavity, capable of shaping the optical local density of states. Over the past three decades, optical cavities have been used in a plethora of applications encompassing quantum technologies, optoelectronics, photonics, and spectroscopy~\cite{lanzillotti-kimura_enhanced_2011-1, Peter_PRL, somaschi_near-optimal_2016, amo_superfluidity_2009,st-jean_lasing_2017, lateral_Bragg_Papatry, Data_Comm_L3Matrix, skolnick_strong_1998-1, amo_polariton_2011, Ducci_2017,QD-single-photon1, QD-single-photon2}. Similarly, acoustic cavities utilizing the same DBR configurations can confine and enhance phononic fields, offering prospects for ultra-high frequency applications~\cite{trigo_confinement_2002, Huynh_2006, highfreq_phonons}. Furthermore, these cavities were recently used in optomechanics, and as platforms for simulating solid-state physics phenomena~\cite{priya_2023, Esmann_Optica_2019, ortiz_topological_2021, rodriguez_topological_2023, chafatinos_asynchronous_2023-1, chafatinos_polariton-driven_2020, anguiano_micropillar_2017, kent_acoustic_2006, czerniuk_lasing_2014}.
 
The widespread use of DBRs and their resonant structures emphasizes the importance of developing accurate tools for optimized device design. Current designs and evaluations of optical and nanophononic devices often focus solely on the number of DBR periods and materials~\cite{Kimura_2007}. However, the reality of material growth and fabrication always entails some surface roughness, even in samples produced using cutting-edge techniques. While the role of interface roughness has been discussed in some cases~\cite{Rozas_2009_PRL}, the impact of surface thickness variation and surface roughness is commonly overlooked, presumably due to the assumption that its influence might be negligible. To justify this conjecture, one would have to assume surface roughness of negligible magnitude, combined with the fact that the confined mode is mainly localized within the cavity spacer. However, in reality, some variations of the top layer thickness often occur during device processing, or due to oxidation of the top layer of structure. Additionally, some degree of surface roughness is invariably present even when using state-of-the-art growth techniques~\cite{RefInd,Qdash,Wang_2018}. The resulting top layer variations and roughness can significantly affect the performance of DBR resonators, which rely on precise definition of the layered structure to define the resonant frequencies. 

In this study, we delve into the impact of top layer thickness and surface roughness on the phonon dynamics and quality factor (Q-factor) in Fabry-Perot (FP) cavities and nanoacoustic open resonators based on DBRs. Our investigation centers on GaAs/AlAs DBR superlattices, which are extensively utilized in both acoustic and optical micro-cavities. We examine both Fabry-Perot and open-cavity resonators, unraveling the influence that even small variations in the last layer thickness and roughness can exert on the Q-factor, which is the metric that quantifies energy dissipation in these resonators. 

Moreover, open resonators are particularly useful for applications that demand environmental responsivity. What is more, additional materials can be deposited on them to form an acoustic resonator responsive to external stimuli. We explore practical scenarios where these cavities function as sensors with dielectric materials positioned atop them, allowing for a direct assessment of the impact of roughness on device performance. Specifically, we incorporate VO$_{2}$ and mesoporous SiO$_{2}$ materials, conducting a detailed exploration of the parameter space to assess the combined influence of roughness and number of DBR periods on the quality factor. Materials sensitive to external stimuli, such as mesoporous thin films reacting to humidity changes~\cite{ruminski_humidity-compensating_2008,cardozo_de_oliveira_design_2023,boissiere_porosity_2005-1}, and VO$_{2}$ responding to thermally or optically induced phase transitions~\cite{qazilbash_mott_2007,cavalleri_femtosecond_2001}, undergo changes in their elastic properties under such stimuli~\cite{dong_elastic_2013,benetti_photoacoustic_2018}. For instance, mesoporous thin films are known to adsorb liquids into their pores, and changes in the relative humidity lead to water condensation inside the pores, which alters their mechanical properties. On the other hand, VO$_{2}$ exhibits two phases -monoclinic and tetragonal-, and phase transitions can modify their mechanical properties. These changes influence the speed of sound, affecting the response of such resonators. Such materials and designs could potentially impact responsive nanoacoustic and nanophotonic devices.

In Section \ref{sct:Model_Implementation}, we analyze the main sources and types of roughness and thickness variation in the top layer of GaAs/AlAs structures, and present the model that we employed to analyze their impact. The initial part of the study focuses on examining the influence of the top layer thickness on the responses of both Fabry-Perot resonators and open cavities, as elaborated in Section \ref{sct:top-layer-dependence}. Building upon this analysis, Section \ref{sct:roughness} delves into quantifying the impact of layer thickness and surface roughness on these resonators. In Sections \ref{sct:external-stimuli} and \ref{sct:enhance_q-factor}, we introduce the dielectric materials mesoporous SiO$_{2}$ and VO$_{2}$ on top of a DBR and explore the resonator response as a function of their roughness. Within these sections, we also quantify the effectiveness of methods used to mitigate the effects of roughness, such as polishing or planarization. 

\section{Model Implementation} \label{sct:Model_Implementation}

To simulate the structure, we employ a model based on the transfer matrix method (TMM). We consider free-strain boundary conditions and solve the standard 1D wave equation. Detailed implementation specifics of the TMM for studying transmission and reflectivity spectra in multi-layered superlattices are outlined in~\cite{fainstein_raman_2007,lanzillotti-kimura_phonon_2007}. We here extend this method to account for surface roughness, introducing a simplified model that simulates the effects of roughness on acoustic FP and open cavity resonators. Surface roughness can have various sources, resulting in either more (additive roughness) or less (subtractive roughness) material than originally designed. In GaAs/AlAs structures additive or subtractive roughness may arise during material growth or device processing, and as a result, it can vary in terms of its type and direction. 

For example, a relatively small variation on the order of one monolayer (ML) ($\sim$ $ 0.5 \: nm$), is commonly observed in high-quality epitaxially grown flat layers, such as MBE-grown GaAs/AlAs or InGaAs/InP structures~\cite{RefInd,Qdash,panish_molecular_1980}. This roughness tends to be discrete, usually resulting in 1-ML-thick plateaus in the lateral direction. Non-flat layers, like quantum dot (QD) nanostructures, frequently integrated within DBRs due to their exceptional optical and quantum properties, might contribute to nm-scale roughness in the layers grown on top of them. Larger roughness is expected when an additional layer is placed atop the MBE-grown cavity, or for samples grown using different techniques. In such cases, the roughness can vary significantly based on the specific technique and material, typically ranging from a few nanometers to a few tens of nanometers. Furthermore, additional roughness or width variations of the top layer might arise due to oxidation~\cite{RefInd}, or during device fabrication processes, such as etching or mask removal steps. This roughness' texture is random and continuous in the vertical direction, and it varies from sample to sample.

Given the diversity of roughness types, and the practical impossibility of knowing its precise structure in each sample, developing a full atomic-scale model for roughness becomes highly impractical. For this reason, in this work we have developed and implemented a model that approximates the roughness of one sample as a set of samples that exhibit planar surfaces and different thicknesses, averaging their simulated surface displacements. Specifically, we model the surface roughness by considering a normal distribution of thickness variation for the top layer of the structure, with a standard deviation $\sigma$. For each case, we average the resulting spectra over 2000 iterations of random thickness distributions with a specific $\sigma$, while keeping the parameters of the other layers of the DBR constant. Subsequently, we fit the resonant modes using a Gaussian function and calculate the Q-factor by computing the ratio between the resonant frequency and the linewidth extracted from the fitting. Unless explicitly stated differently, we imply this averaging process when we refer to roughness simulation in this paper. 

In the outlined approach, we work under the assumption that the inhomogeneous broadening approximation remains valid. Specifically, surface roughness introduces inhomogeneous broadening of the resonant modes, which influences the resonator's response and compromises its Q-factor. The studied structures are based on GaAs/AlAs DBR superlattices with flat interfaces, finalized with the top layer which contains a rough surface. The periodicity of the superlattice introduces Brillouin zone folding and miniband openings at frequencies $\Omega_{m}=m\pi v/d$, where $m\in\mathbb{N}$, and $d$ and $v$ represent the unit cell thickness and phonon group velocity, respectively~\cite{phonon_folding_1,phonon_folding_2, highfreq_phonons}. Throughout our analysis, we consider appropriate thickness variations and roughness ranges, taking into account the typical values observed in practical samples. The material parameters for GaAs, AlAs, mesoporous SiO$_{2}$, and VO$_{2}$, utilized in our analysis, are drawn from relevant literature sources~\cite{RefInd,adachi_gaas_1985,abdala_mesoporous_2020,dong_elastic_2013}.

\section{Influence of Top Layer Thickness on Fabry-Perot and Open Resonators} \label{sct:top-layer-dependence}

\begin{figure*}[!ht]
\centering
\includegraphics[scale = 0.35]{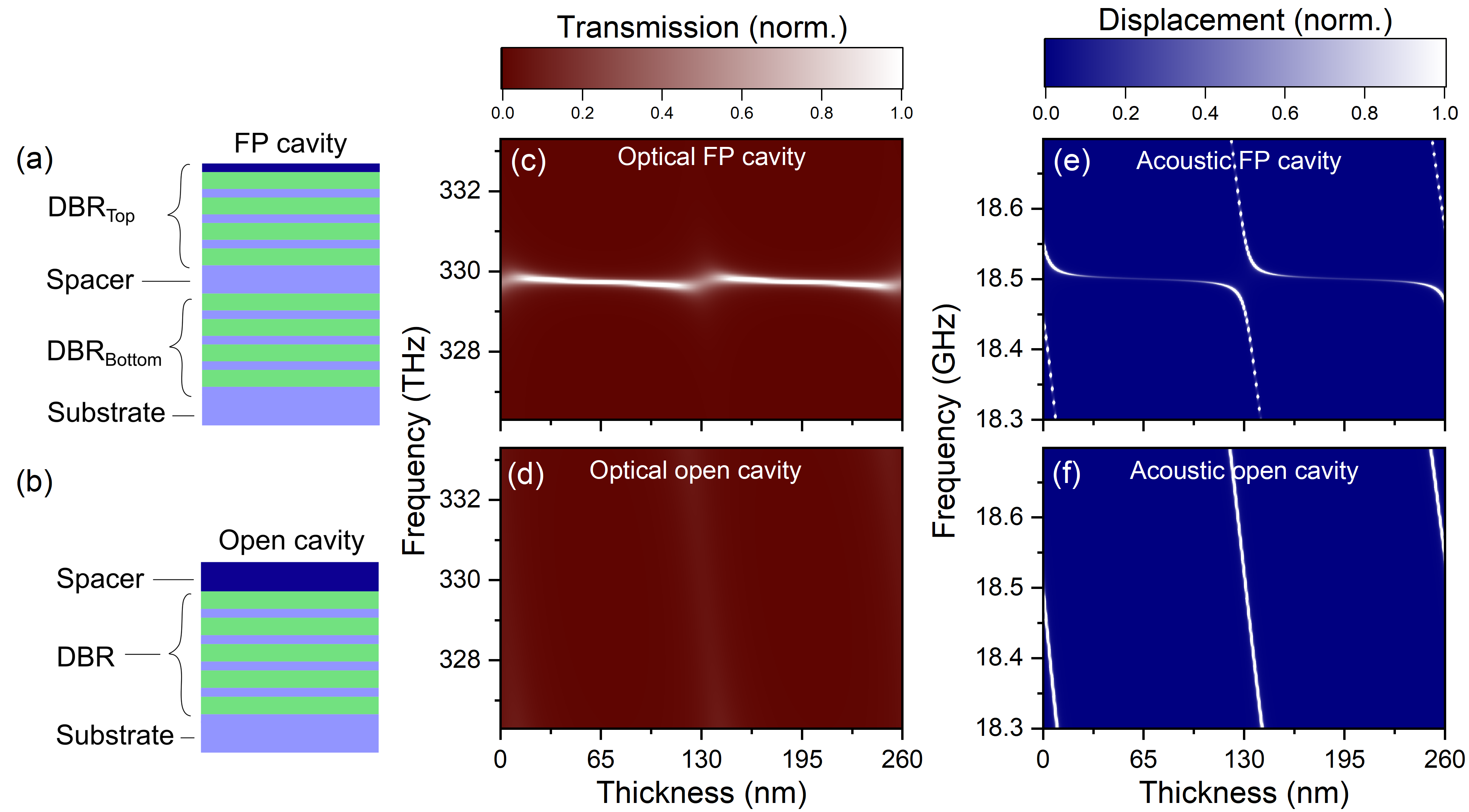}
\caption{Effects of top layer thickness variation on Fabry-Perot and open-cavity resonators. (a,b) Schematics of (a) Fabry-Perot resonator with a 16-period top DBR and a 20-period bottom DBR and (b) open cavity with 20 DBR periods and $\lambda/2$ spacer. (c,d) Colormaps of the optical reflectivity as a function of the top layer thickness for (c) Fabry-Perot and (d) open cavity. (e,f) Colormaps of the acoustic displacement as a function of the top layer thickness for (e) Fabry-Perot and (f) open cavity.}
\label{fig:FP_cavity}
\end{figure*}

Our study begins with an analysis of how variations in the top layer thickness impact the response of Fabry-Perot and open resonators. These fluctuations in top layer thickness are common in practical samples and can arise due to  processes like oxidation of the top layer upon exposure to the atmosphere~\cite{RefInd}, or during device fabrication steps such as etching or mask removal ~\cite{lateral_Bragg_Papatry}. Schematics of the investigated resonators are presented in Figure~\ref{fig:FP_cavity}(a,b). 

The first structure (Fig.~\ref{fig:FP_cavity}(a)), is a conventional FP cavity, incorporating a $\lambda/2$-cavity between two GaAs/AlAs DBR superlattices, with $\lambda$ representing the acoustic wavelength. This configuration results in a high-quality cavity for both photons and phonons due to the nearly identical refractive index contrast and acoustic impedance contrast of GaAs and AlAs~\cite{fainstein_strong_2013}. Moreover, the closely matched lattice constants of these materials facilitate the growth of thick, high-quality layers using standard epitaxial techniques, making them ideal for applications involving high Q-factor microcavities.  

The second design under investigation is the open-cavity resonator (Fig.~\ref{fig:FP_cavity}(b)), formed by a $3\lambda/2$ low-acoustic-loss spacer cavity on top of a single DBR. In the absence of surface roughness, in this arrangement, the free surface acts as a perfect mirror for phonons at the studied frequencies, since they cannot propagate toward the air. Consequently, when combined with a high-reflection bottom DBR, this arrangement generates a cavity with a high Q-factor, rendering it an ideal platform for our investigation. 

We performed a comprehensive study of the parameter space to assess the combined effects of the top layer thickness and DBR periods on the quality factor. In all configurations, we maintain a constant difference of 4 GaAs/AlAs layer pairs between the top and bottom DBRs. This difference ensures a symmetric optical cavity due to the higher refractive index contrast at the DBR/air interface compared to the DBR/substrate interface. Figure~\ref{fig:FP_cavity}(c) and (d) present colormaps of the optical reflectivity as a function of the top layer thickness, illustrating the dependence on the top layer thickness for both FP and open cavities, respectively. Figure~\ref{fig:FP_cavity}(c) shows a minimal sensitivity of the FP frequency over the whole thickness range except for $\lambda/2$ which leads to a small frequency shift. In the open cavity (Figure~\ref{fig:FP_cavity}(d)), we observe a sharp frequency shift for a small thickness change around $\lambda/2$, corresponding to a slope of ~0.34 THz/nm. Additionally, the transmission is much weaker than the FP case because the optical quality factor is lower in the open cavity due to the missing upper high-reflectivity DBR.

Figures~\ref{fig:FP_cavity}(e) and (f) depict colormaps of the acoustic displacement for a cavity with the same DBRs as Fig.~\ref{fig:FP_cavity}(c) and (d). Notably, as shown in Fig.~\ref{fig:FP_cavity}(e), the FP cavity exhibits a range between approximately 30 and 100 nm in which the mode frequency demonstrates minimal sensitivity to the thickness change, featuring a slope of $\sim0.1$ MHz/nm. This linear regime over a broad thickness range results in a frequency shift of the acoustic resonance of less than 10 MHz. Conversely, the open cavity exhibits a strong dependence of the acoustic resonance on the top layer thickness, with the acoustic mode frequency inversely proportional to the layer thickness. Around the midpoint of $\lambda/2$ this curve exhibits a slope of $\sim18$ MHz/nm, significantly larger than the FP cavity case. In this regime, small fluctuations in the thickness of a rough open cavity are expected to induce a pronounced broadening of the acoustic resonance, which we will investigate further in the subsequent section. 

Comparing optical and acoustic resonators, we observe two main differences. The first one is the anti-crossing observed in the acoustic Fabry-Perot (FP) resonator at 18.5 GHz (Fig.\ref{fig:FP_cavity}(e)), which is not evident in the optical FP (Fig.\ref{fig:FP_cavity}(c)). At a thickness of $\lambda/2$, a second cavity due to reflection at the surface is formed in both cases. However, the acoustic cavity is much stronger, thus lifting the degeneracy of the modes. The second difference concerns the open cavity, where the transmission for the optical mode (Fig.\ref{fig:FP_cavity}(d)) is much weaker than for the acoustic mode (Fig.\ref{fig:FP_cavity}(f)). These differences arise from the fact that the free surface acts as a perfect mirror for acoustic phonons, whereas this is not the case for photons, as the electromagnetic field can penetrate into the air, creating a much higher quality factor cavity for phonons compared to photons. 

\section{Influence of Surface Roughness on the Quality Factor of Fabry-Perot and Open Resonators} \label{sct:roughness}

\begin{figure}[!h]
\centering
\includegraphics[scale = 0.35]{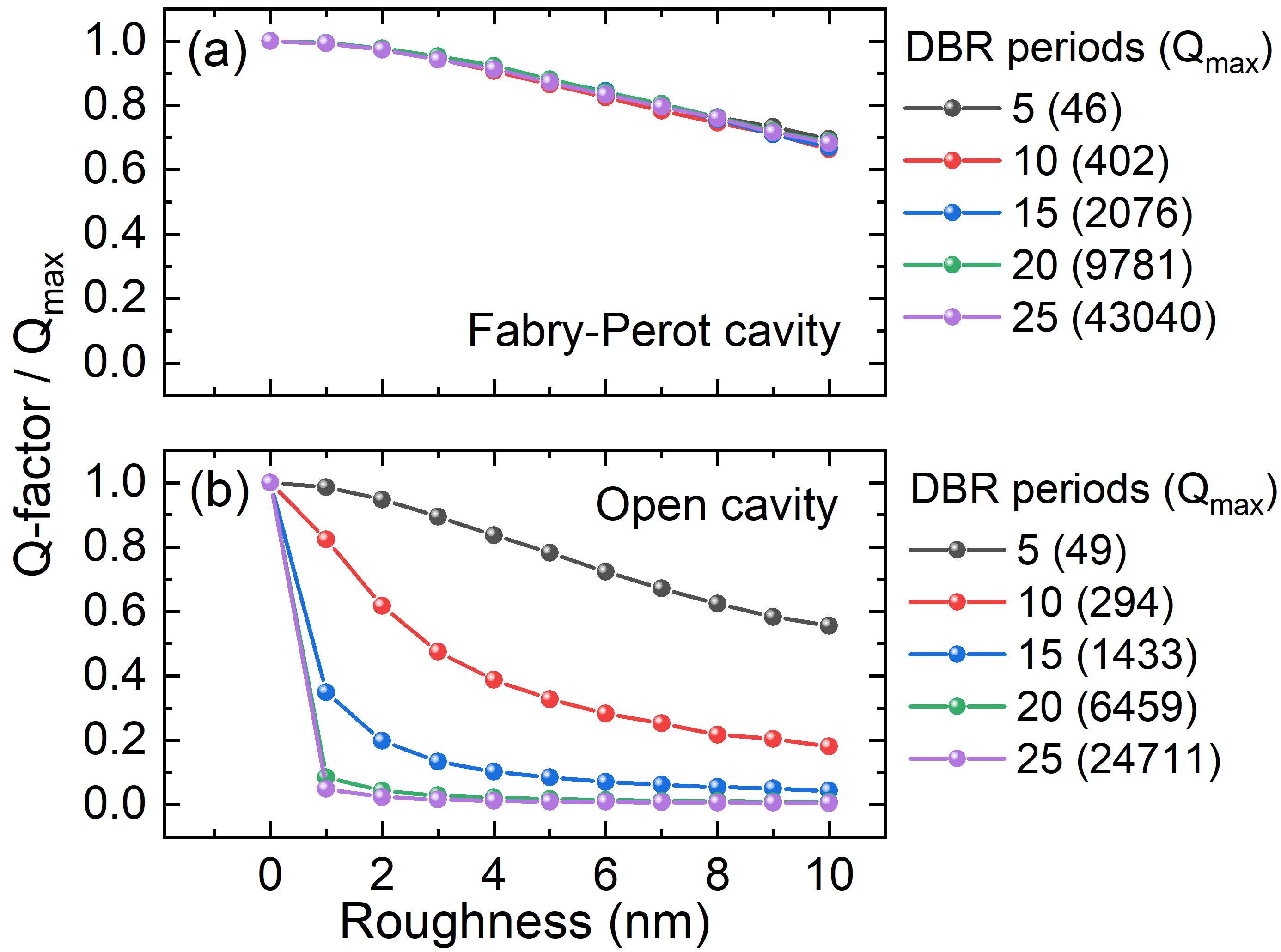}
\caption{Effects of roughness on the Q-factor of (a) a $\lambda/2$ cavity Fabry-Perot resonator, and (b) a 3$\lambda/2$ open-cavity resonator for various DBR configurations. In both panels, the bottom DBR varies from 5 to 25 bilayers in intervals of 5, while in (a) the top DBR of the Fabry-Perot comprises 4 bilayers less than the bottom one in all cases.}

\label{fig:DBR_roughness_Q-factor}
\end{figure}

We now delve into the influence of surface roughness on the Q-factor of both FP and open resonators near their resonant frequencies, considering various DBR configurations. 
Figure~\ref{fig:DBR_roughness_Q-factor}(a) focuses on Fabry-Perot resonators, analyzing five configurations with varying numbers of periods in the bottom DBR (ranging from 5 to 25 in 5-bilayer intervals). In all cases, the top DBR comprises 4 periods less than the bottom DBR (1, 6, 11, 15, and 21). Corresponding results for open cavities are shown in Figure~\ref{fig:DBR_roughness_Q-factor}(b). The two key parameters —roughness and the number of DBR layers— exert a pronounced and predictable influence on the Q-factor.

For structures with flat surfaces ($\sigma=0$~nm), the Q-factor is solely determined by the number of DBR periods. However, as surface roughness increases, the effectiveness of stacking additional layers significantly wanes. This is due to the upper Q-factor limit imposed by increased fluctuations in resonant frequencies, introduced by surface roughness. Although negligible for applications where ultrahigh Q-factors are not a requirement, this effect might have drastic consequences in specific applications, such as polaritonics and single photon sources ~\cite{Peter_PRL, somaschi_near-optimal_2016}. This effect is considerably more prominent in open-cavity resonators than in Fabry-Perot ones, as depicted in Figure ~\ref{fig:DBR_roughness_Q-factor}. In essence, higher roughness dictates a reduced number of DBR periods required to attain the maximum Q-factor in the structure, consequently leading to diminished maximum Q-factor values. 

For instance, an open resonator composed of GaAs-with its characteristic low roughness (approximately one monolayer)-would benefit from an increased number of DBRs. Specifically, assuming a 0-nm-roughness GaAs/AlAs acoustic Bragg mirror consisting of 20 periods achieves a Q-factor of 6459, while 25 periods a Q-factor of 24711, and continually rising as more layers are added. Conversely, 1-nm-roughness would result in a Q-factor of 476 for 20 DBR layers and 505 for 25 DBR layers or more (Q-factor saturates above 25 layers), demonstrating a significant reduction of the Q-factor even for such minor roughness. Additionally, when the same resonator possesses a 2 nm roughness on its top layer, or a material with this roughness is deposited on top of it, the Q-factor saturates already at 20 DBR layers and adding more layers fails to increase it, as indicated in Fig.~\ref{fig:DBR_roughness_Q-factor}(b).  

A comparison between the FP and open cavity configurations, reveals that the open cavity Q-factor is more sensitive to surface roughness than the FP cavity. Notably, even under relatively high surface roughness conditions (10 nm), the Q-factor of the FP cavity does not saturate; although it decreases to approximately 50 $\%$ of its maximum value. Overall, our study suggests that factoring in roughness becomes imperative when engineering high-quality photonic or acoustic resonators. Doing so, would enable precise Q-factor evaluations for each design, and prevent superfluous layer stacking. This, in turn, would improve device design, enhance operational efficiency, and minimize costs. When working with acoustic resonators at higher frequencies, and hence thinner layers, the effect of roughness is extremely critical.

Regarding the DBR unit cell layers' thicknesses, Fig. ~\ref{fig:DBR_roughness_Q-factor} presents results for $\lambda/4-3\lambda/4$ (with respect to the acoustic wavelength $\lambda$). It is worth noting that analogous conclusions hold for $\lambda/4-\lambda/4$ DBRs. Since the overarching findings remain similar, we present here detailed results solely for the former case. We note two differences when the $\lambda/4-\lambda/4$ unit cell is used; first, the stop band becomes twice broader, and second, the mode's resonant frequency exhibits an increased dependency on the top layer thickness. 

\section{Influence of surface roughness on resonators responsive to external stimuli} \label{sct:external-stimuli}

We now proceed to analyze the effects of roughness on the responsivity of GaAs/AlAs DBR-based resonators when an additional top layer of a different material, responsive to external stimuli, is introduced. We focus solely on open-cavity resonators here, due to their anticipated exceptional responsivity to external changes. We simulate resonators with two types of spacers; mesoporous SiO$_{2}$, or VO$_{2}$, as their top layer. Our analysis encompasses three different roughnesses of 0~nm, 2~nm, and 5~nm, and two different phases for each material: 0 and 100$\%$ relative humidity for the mesoporous cavity, and monoclinic and tetragonal phase for the VO$_{2}$ cavity. The selected roughness values of 2-5 nm were based on minimal (optimal) realistic values achievable for optimized SiO$_{2}$ and VO$_{2}$ materials, while the 0~nm case serves as a reference point. In all these cases, the open-cavity DBR consists of 15 GaAs/AlAs pairs. This choice strikes a balance between structure complexity and performance based on the outcomes of Figure~\ref{fig:DBR_roughness_Q-factor}, for the considered roughness values. Specifically, for a 5-nm roughness, stacking additional DBR layers would have a minimal impact on the Q-factor thereby needlessly increasing the complexity of the structure. Conversely, employing fewer layers substantially diminishes the Q-factor. 

\begin{figure}[!h]
\centering
\includegraphics[scale = 0.335]{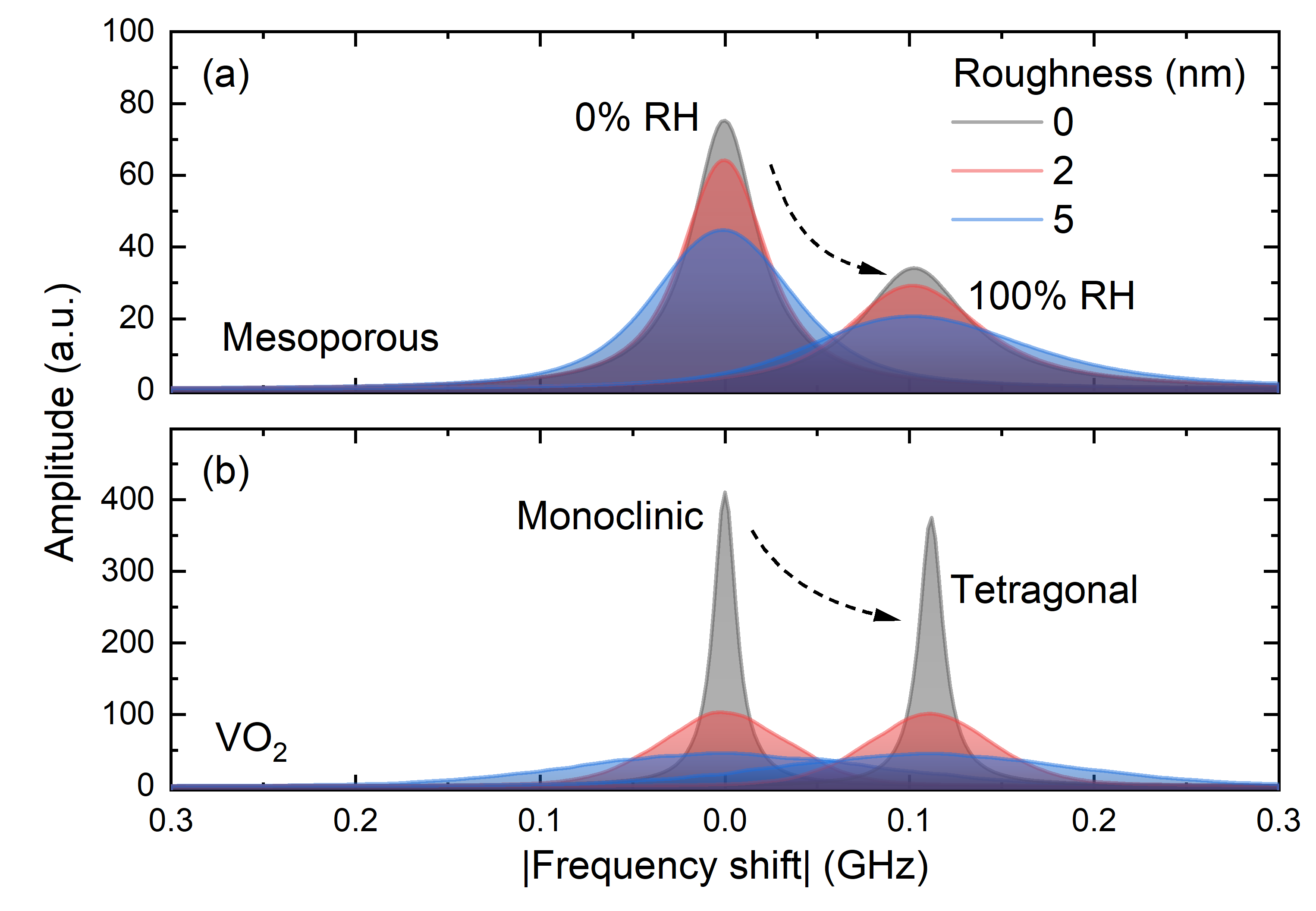}
\caption{Acoustic displacement spectrum of the structure containing 15 GaAs/AlAs DBR periods and a top layer sensitive to external stimuli. The top layers considered are (a) mesoporous SiO$_{2}$ at 0~$\%$ and 100~$\%$ relative humidity, and (b) VO$_{2}$ at the monoclinic and tetragonal phases. A roughness of $\sigma = 0$, $2$, and $5$~nm is considered in both cases.}
\label{fig:meso_vo2_shift}
\end{figure}

We simulate the humidity in the mesoporous material by considering a weighted average for the material density and speed of sound, taken from a combination of dense SiO$_{2}$, air, and water. As for VO$_{2}$, the acoustic properties in its two phases are drawn from Reference ~\cite{dong_elastic_2013}. Central findings are illustrated Figure~\ref{fig:meso_vo2_shift}, displaying the phonon displacement spectrum of the simulated structures. Figure~\ref{fig:meso_vo2_shift}(a) focuses on the mesoporous material, comparing its response to 0~$\%$ and 100~$\%$ relative humidity, for three different surface roughness values. As demonstrated, the cavity's response to humidity change is evident, with peak amplitude reduction and frequency shift as humidity rises~\cite{cardozo_de_oliveira_probing_2023,cardozo_de_oliveira_design_2023}. However, increased roughness leads to broader peaks, potentially making frequency shifts harder to resolve. For a 5~nm roughness, the largest considered, the shift remains resolvable. Nonetheless, the trend suggests that roughness exceeding 5~nm -realistic for non-optimized SiO$_{2}$ samples- or smaller humidity changes, could render the system response insufficient. 

Figure~\ref{fig:meso_vo2_shift}(b) reveals changes in the acoustic displacement spectra of VO$_{2}$ across its two phases, given the same three surface roughness values. Results show that minor roughness —around 2~nm or less— yields well-defined peaks, making phase changes clearly resolvable. Conversely, roughness around 5~nm leads to larger broadening, affecting the resolution of frequency shifts. The broadening here is especially pronounced compared to the mesoporous case. This is due to the speed of sound being considerably larger in  VO$_{2}$ compared to the mesoporous material. This means that a similar thickness change causes a larger frequency shift in VO$_{2}$, thus enhancing the roughness-induced inhomogeneous broadening. These outcomes indicate that surface roughness considerably impacts Q-factors of GaAs/AlAs DBR open cavities utilized for sensing. Surface roughness reduces the system's ability to resolve external changes, even for modest roughness levels, with similar or larger values expected in realistic samples. The subsequent section will explore methods to mitigate roughness impact and enhance device sensitivity.

\section{Strategy to enhance Q-factor} \label{sct:enhance_q-factor}

As we have seen in the preceding sections, surface roughness can significantly reduce the quality factor of the cavity, and consequently its sensitivity to external changes. Although Section \ref{sct:external-stimuli} showed that open-cavity resonators with rough surfaces can still resolve frequency changes induced by external stimuli, the effective reduction in Q-factor compared to an ideal no-roughness scenario remains substantial. It would, therefore be highly desirable to minimize the roughness as much as possible in practical structures, especially in those designed for sensing applications.~\cite{cardozo_de_oliveira_design_2023} The "flattening" of rough surfaces can be achieved in various ways, most commonly via either polishing the surface or by depositing additional composite thin films. The latter is usually a much simpler approach, and it is preferred when adding such a layer does not have a negative effect on the other properties of the sample, while the former is preferred otherwise. 

\begin{figure}[!ht]
\centering
\includegraphics[scale = 0.38]{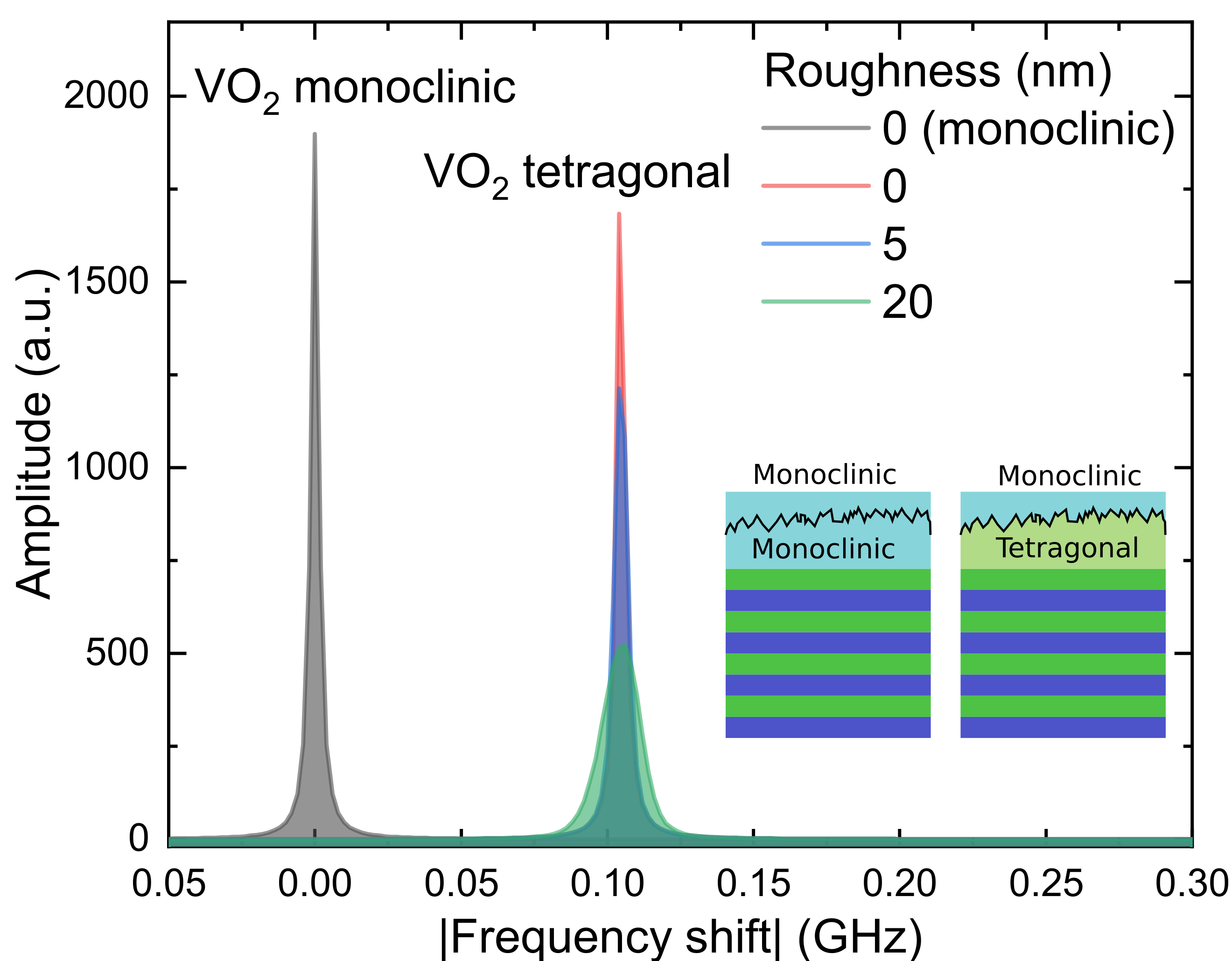}
\caption{Acoustic displacement spectrum of an open cavity with 20 GaAs/AlAs DBR periods, finalized with a rough VO$_{2}$ and a flat capping layer containing the elastic properties of VO$_{2}$ in the monoclinic phase. The black peak corresponds to the displacement spectrum of monoclinic VO$_{2}$, whereas the red, blue, and green peaks, to the tetragonal VO$_{2}$ with 0-nm, 5-nm, and 20-nm roughness, respectively.}
\label{fig:VO2_capping_layer}
\end{figure}

The polishing method removes the top rough part of the material by either rubbing it or by applying a chemical treatment resulting in a smooth surface, while the planarization method deposits a polymer resist on the top of the sample. As the polymer composite is spin-coated on the sample in liquid form, it tends to even out the surface roughness, before it is cured into a solid polymer with a smooth surface. We have studied the effect of such surface-reduction methods on our structure, and the results are summarized in Figure~\ref{fig:VO2_capping_layer}. To account for both methods, we have simulated two similar structures that both have an additional layer placed atop. We simulate the planarization method as follows: first, we consider a rough layer of 243.24 nm of monoclinic VO$_{2}$. Then, we add a 20-nm-thick ($d_{2}$) layer of a virtual material. The roughness of the two top layers is complementary in shape in such a way that the total thickness $D$ remains constant at 263.24 nm, without any surface roughness. The added virtual material has the same elastic parameters as the monoclinic VO$_{2}$, as shown in the left schematic of Fig.~\ref{fig:VO2_capping_layer}. The acoustic resonance corresponds to a monoclinic-VO$_{2}$ $\lambda/2$ cavity. In the second case, we assume the same structure as the previous case, but with the bottom layer in the tetragonal VO$_{2}$ phase (different mass density and speed of sound), as illustrated in the right schematic of Fig.~\ref{fig:VO2_capping_layer}).

Figure~\ref{fig:VO2_capping_layer} illustrates the results for both material combinations, for different values of original roughness (i.e. before adding a planarization layer). Our original structures consisted of 20 GaAs/AlAs DBR periods with a thin VO$_{2}$ layer on top. In the case of 0-nm-roughness, both peaks have relatively high Q-factors. The structure with bottom and top layers with VO${2}$ in the monoclinic phase is entirely insensitive to roughness as both layers have the same acoustic properties. For the other structure, as the simulated roughness progressively increases to 5~nm and 20~nm, the quality factor decreases. Even though the top surface is flat, there is still an effective roughness remaining in the second structure, in-between the last two layers, coming from the fact that the initial VO$_{2}$ layer was rough, and the second layer had different properties. However, there is still a clear improvement compared to the case when the additional layer is not deposited (see Fig~\ref{fig:meso_vo2_shift}). The acoustic mode associated with the monoclinic phase presents an improved quality factor due to the absence of surface roughness. In the case of the tetragonal phase, the impedance matching between the air and the two layers also increases the quality factor of the resonator with respect to the case where the roughness is directly present at the surface of the device.

\section{Conclusions}
In this study, we conducted an analysis of the impact of top layer thickness and surface roughness on the quality factor and performance of acoustic and optical resonators based on GaAs/AlAs DBRs. Our findings reveal that thickness inaccuracies appreciably influence the mode frequency, while even relatively small surface roughness, on the order of a few nanometers, significantly influences the quality factor. Consequently, these effects can have a substantial impact on the device performance of DBR microcavities. This important aspect is often overlooked in the design process, making it essential to consider surface roughness for optimal performance. Notably, our developed model allows us to determine the maximum achievable quality factor and the minimum number of DBR layers required to achieve it, for different surface roughness values. By doing so, we can eliminate the need for growing unnecessary additional layers that do not contribute effectively, leading to more efficient and cost-effective designs. 

We observed that surface roughness affects acoustic cavities more prominently than their optical counterparts, although both are significantly influenced. Additionally, open-cavity designs are much more sensitive to roughness compared to FP cavities, which is expected due to the closer proximity of roughness to the cavity region, leading to a more pronounced interaction. Moreover, we explored practical case studies in which such microcavities were designed to serve as sensors. Our investigations provided valuable insights as to how surface roughness impacts these applications, and enabled us to assess strategies for mitigating its effects. We believe that the insights obtained in this study have broad implications for the design of efficient optoelectronic, photonic, and acoustic devices.

\section{Acknowledgments}
 The authors acknowledge funding from European Research Council Consolidator Grant No.101045089 (T-Recs). This work was supported by the European Commission in the form of the H2020 FET Proactive project No. 824140 (TOCHA).

\bibliography{references2}

\end{document}